\documentclass[12pt]{article}
\usepackage[dvips]{graphicx}
\usepackage{pdproc}

\usepackage{graphicx}% Include figure files
\usepackage{dcolumn}% Align table columns on decimal point
\usepackage{bm}% bold math
\usepackage{amssymb}  
\usepackage{amsmath}
\usepackage{epsfig}    

\newcommand{\tr}{\mbox{tr}}
\def\beq{\begin{equation}}
\def\eeq{\end{equation}}
\def\bea{\begin{array}}
\def\eea{\end{array}}
\def\be{\begin{equation}}
\def\ee{\end{equation}}
\def\ba{\begin{eqnarray}}
\def\ea{\end{eqnarray}}

\def\to{\rightarrow}

\def\[{\left[}
\def\]{\right]}
\def\({\left(}
\def\){\right)}

\def\sm0{{\widetilde{m}_0}}

\def\U1em{{U(1)_{\rm em}}}
\def\to{\rightarrow}

\def\sq2{\sqrt{2}}

\def\ee{e^+e^-}

\def\End{\end{document}}
\def\Journal#1#2#3#4{{#1} {\bf #2}, #3 (#4)}
\def\NPB{{\rm Nucl. Phys.} B}
\def\PLB{{\rm Phys. Lett.}  B}

\def\PRD{{\rm Phys. Rev.} D}

\def\fsl#1{\setbox0=\hbox{$#1$}                 % set a box for #1 
   \dimen0=\wd0                                 % and get its size
   \setbox1=\hbox{/} \dimen1=\wd1               % get size of /
   \ifdim\dimen0>\dimen1                        % #1 is bigger
      \rlap{\hbox to \dimen0{\hfil/\hfil}}      % so center / in box
      #1                                        % and print #1
   \else                                        % / is bigger
      \rlap{\hbox to \dimen1{\hfil$#1$\hfil}}   % so center #1
      /                                         % and print /
   \fi}  
\newcommand{\VEV}[1]{\langle #1 \rangle}

  \makeatletter 
  \def\@cite#1{[#1]} 
  \makeatother    
\begin{document}

\renewcommand{\thefootnote}{\alph{footnote}}

\title{
A Mechanism for the Top-Bottom Mass Hierarchy
}

\author{MICHIO HASHIMOTO\,$^1$, and SHINYA KANEMURA\,$^2$}

\address{ 
$^1$Department of Physics,  
Pusan National University, Pusan 609-735, Korea\\
$^2$Department of Physics, 
Osaka University, Toyonaka, Osaka 560-0043, Japan
\\ 
{\rm $^1$michioh@charm.phys.pusan.ac.kr} \\
{\rm $^2$kanemu@het.phys.sci.osaka-u.ac.jp}} 

\abstract{
We discuss a mechanism to generate hierarchy 
between masses of the top and bottom quarks 
without fine-tuning of the Yukawa coupling constants.
In the framework of the two-Higgs-doublet model (THDM) 
with a discrete $Z_2$ symmetry, there exists 
the vacuum where only the top quark receives the mass of 
the order of the electroweak symmetry breaking scale 
$v(\simeq 246\,\mbox{GeV})$, 
while the bottom quark remains massless. 
We show a model in which a small sofb-breaking mass 
$m_3^2 [\sim v^2/(4\pi)^2]$ for the $Z_2$ symmetry 
is generated by the dynamics above the cutoff scale of the THDM.
The appearance of $m_3^2$ gives the small mass for the 
bottom quark, and explain the hierarchy $m_b/m_t \ll 1$.
}

\normalsize\baselineskip=15pt

\section{Introduction}

The measured quark mass spectrum shows a specific feature. 
Only the top quark has the mass of the order of the 
electroweak symmetry breaking (EWSB) scale $v$ 
($= (\sqrt{2} G_F^{})^{-1/2} \simeq 246$ GeV), while masses of the 
other quarks are much smaller. In the Standard Model (SM), 
the unique Higgs doublet field $\Phi_{\rm SM}^{}$ is
responsible for the EWSB and gives masses of all quarks 
via the Yukawa interactions; i.e.,   
$m_f \simeq y_f \langle \Phi_{\rm SM}^{} \rangle$ with
$\langle \Phi_{\rm SM}^{} \rangle = (0, v/\sqrt{2})^T \!$.
The observed mass spectrum is obtained only by 
assuming unnatural hierarchy among the Yukawa coupling constants $y_f$. 
Nevertheless, no explanation is given for such fine tuning in the SM. 

In this Talk, we present an alternative scenario in which 
the quark mass spectrum is reproduced without fine tuning 
in magnitude of the Yukawa coupling constants\cite{hk}. 
We study the hierarchy between $m_t$ and $m_b$ 
under the assumption of $y_t \sim y_b \sim {\cal O}(1)$. 
In order to realize $m_b/m_t \sim 1/40$ in a natural way,  
we consider the two-Higgs-doublet model (THDM) with 
$\Phi_1$ and $\Phi_2$, imposing the discrete $Z_2$ symmetry\cite{z2},  
in which only $\Phi_1$ couples to the bottom quark while 
$\Phi_2$ does to the top quark.
The hierarchy $m_t \gg m_b$ is then equivalent to
$v_2 \gg v_1$, where 
$\langle \Phi_{1,2} \rangle = (0, v_{1,2}/\sqrt{2})^T$.
When the $Z_2$ symmetry is exact, there exists the vacuum with
$v_1 = 0$ and $v_2 = v$. A nonzero value of $v_1$ $(\ll v_2)$ 
is induced as a perturbation of a small soft-breaking parameter 
$m_3^2$ for the $Z_2$ symmetry.
The small $m_3^2 [\sim v^2/(4\pi)^2]$ is generated by
the dynamics above the cutoff scale of the THDM.
Consequently, we obtain $m_b/m_t \ll 1$.
This scenario can be extended to include the first two generation 
quarks. 

\vspace*{-2mm}
\section{Minimal Model} 

We first consider only the top and bottom quarks among fermions, and 
discuss the extension for the other quarks later on. 
Details are shown in Ref.~\cite{hk}.
In order to describe the assumption of $y_t \simeq y_b$, we 
introduce the global $SU(2)_R^{}$ symmetry\cite{cs,gs_2hdm}, 
in addition to the $SU(2)_L$ gauge symmetry:

\vspace*{-2mm}
\noindent
\begin{eqnarray}
  q_{L,R}^{}  \to  q_{L,R}' = U_{L,R}^{} \; q_{L,R}^{}\, , 
  M_{21}^{}   \to  M_{21}' = U_L M_{21} U_R^\dagger, 
\end{eqnarray}

\vspace*{-2mm}
\noindent
where $q_{L,R}^{} \equiv (t_{L,R}^{},b_{L,R}^{})$ and 
$U_{L,R} \in SU(2)_{L,R}$, respectively.
The $2 \times 2$ matrix $M_{21}^{}$ is defined by 
$M_{21} \equiv \left( \tilde{\Phi}_2, \Phi_1 \right), \;\;\;
{\rm with} \;\;\;
\tilde{\Phi}_2 = i\tau_2 \Phi_2^\ast$. 
The $Z_2$ symmetry can be expressed in terms of $q_{L,R}^{}$ 
and $M_{21}^{}$ by
$  q_L        \to  q'_L = q_L , \;\;\;
  q_R \to  q'_R = \tau_3 q_R , 
  M_{21}^{}  \to  M_{21}' = M_{21} \tau_3$.
The Yukawa interaction then is written as
${\cal L}_Y^{} =
- y \bar{q}_L^{} M_{21}^{} q_R^{} 
 + \mbox{(h.c.)}\, , \label{L_Y}
$ with $y \equiv y_t = y_b$. 
The Higgs potential 
with the softly-broken $Z_2$ symmetry is given by

\vspace*{-2mm}
\noindent
\begin{eqnarray}
V(M_{21}^{}) &=& 
 \frac{1}{2}m^2 \tr(M_{21}^\dagger M_{21})
       -\frac{1}{2} \Delta_{12}\tr(M_{21}^\dagger M_{21}\tau_3)
      - \left[ m_3^2 \,\det M_{21} + \mbox{(h.c.)}\,\right]
       \nonumber \\ && 
\hspace*{-1.4cm} 
       + \lambda\left[\,\tr(M_{21}^\dagger M_{21})\,\right]^2
%       \nonumber \\ && 
%\hspace*{-1.4cm} 
       + 2\lambda_4 \det (M_{21}^\dagger M_{21})
       + \left[\, \lambda_5 (\det M_{21})^2+ \mbox{(h.c.)}\,\right]. 
   \nonumber
    \label{pot_M} 
\end{eqnarray}

\vspace*{-2mm}
\noindent
The $Z_2$ symmetry is softly broken 
by the mass term of $m_3^2$.  
A non-zero value of $\Delta_{12}$
measures the soft breaking of the global $SU(2)_R^{}$ symmetry. 
In order to evade explicit CP violation,
we choose the phases in $m_3^2$ and $\lambda_5$ 
to be zero. 

Let us consider the effective potential $V(\VEV{M_{21}})$
to study the vacuum structure. 
By using $SU(2)_L$ and $U(1)_Y$, 
the VEV's in the THDM can be generally 
parameterized as 

\vspace*{-2mm}
\noindent
\begin{eqnarray}
 \langle M_{21} \rangle =  \frac{1}{\sqrt{2}}
   \left( \begin{array}{cc} 
              v_2 & v_E^{}   \\
              0   & v_1 + i v_A^{} \\ \end{array}  \right). \label{M21}  
\end{eqnarray}

\begin{figure}[t]
\begin{center}
\includegraphics*[width=6cm]{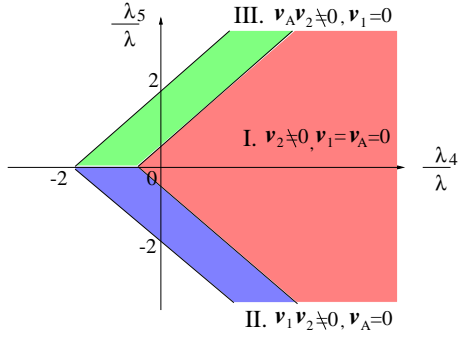}
\caption{%
Vacuum structure for $m_3^2=0$ and $m_2^2 < -|m_1^2|$.}
\label{fig:phase}
\end{center}
\end{figure}

\vspace*{-2mm}
\noindent
In our model, it is easily shown that $v_E^{}=0$ at the tree level.
Since spontaneous CP violation does not occur for $m_3^2 = 0$,
three types of the nontrivial vacuum are possible at the tree level:
(a) $v_1 = v_A = 0, v_2 \ne 0$,
(b) $v_1 v_2\neq 0, v_A=0$,
(c) $v_A v_2\neq 0, v_1=0$.
In Fig.~\ref{fig:phase}, the area (I) corresponds to the vacuum (a), 
while the areas (II) and (III) do to the vacua (b) and (c), respectively.
In order to realize $m_b/m_t \ll 1$ without fine tuning, 
we choose the vacuum (a) which leads to

\vspace*{-2mm}
\noindent
\begin{eqnarray}
 m_t = \frac{1}{\sqrt{2}}\, y \,v,  \quad m_b = 0. 
\end{eqnarray} 

%\begin{figure}[htb]

\vspace*{-2mm}
\indent
We now switch on a {\it small} soft-breaking parameter $m_3^2 (\ll v^2)$ 
of the discrete $Z_2$ symmetry.
We do not consider the possibility of spontaneous CP violation.
A nonzero $v_1$ is necessarily induced for $m_3^2 \ne 0$ from 
the stationary condition.
As a perturbation from the vacuum (a) with $m_3^2 = 0$,
we consequently obtain 

\vspace*{-2mm}
\noindent
\begin{eqnarray}
   \frac{v_1}{v_2}  (\equiv \frac{1}{\tan\beta})  
          &=& \frac{m_3^2}{m_H^2} 
                 \left\{ 
     1 +{\cal O}\left(\frac{m_3^4}{v^4}\right) \right\}.  
\label{ratio}
\end{eqnarray}

\vspace*{-2mm}
\noindent
The masses of the top and bottom quarks are given by 

\vspace*{-2mm}
\noindent
\begin{eqnarray}
 m_t \simeq \frac{1}{\sqrt{2}}\, y \,v, \quad
 m_b = \frac{1}{\sqrt{2}}\, y \,v_1, \label{mt-mb}
\end{eqnarray} 

\vspace*{-2mm}
\noindent
so that the bottom quark finally obtain the small mass. 
The mass hierarchy of $m_t$ and $m_b$ then is deduced from Eqs.~(\ref{ratio})
and (\ref{mt-mb}) without fine tuning of the Yukawa coupling constants; i.e.,
$m_t/m_b = \tan \beta$.
With nonzero $m_3^2$ the Higgs doublets $\Phi_1$ and $\Phi_2$
do mix. The mixing angle $\beta-\alpha$ is expressed as

\vspace*{-1mm}
\noindent
\begin{equation}
 \sin (\beta-\alpha) \!= \! 1 
 - \left(\frac{m_H^2-m_{H^\pm}^2}{m_H^2-m_h^2}\right)^2
     \!\! \frac{2}{\tan^2 \beta}
 + {\cal O}\left(\frac{m_3^6}{v^6}\right), \label{sin-b-a}
\end{equation}

\vspace*{-1mm}
\noindent
From Eq.~(\ref{sin-b-a}), the property of the CP-even Higgs boson 
$h$ is very similar to the SM one. 
The masses of the extra Higgs bosons 
$H$, $A$ and $H^\pm$ turn out to be of the order of $v$. 

We now give comments on the case with relaxed $SU(2)_R$.
First, it can be shown that Eq.~(\ref{ratio}) does not change.
Second, although Eq.~(\ref{sin-b-a}) is slightly modified, 
the essential result of $\sin (\beta-\alpha)=1-{\cal O}(\tan^{-2}\beta)$ 
still holds. Also, the masses of the extra Higgs bosons 
remain to be of ${\cal O}(v)$.

\vspace*{-2mm}
\section{A mechanism for small $m_3^2$.}

Let us consider a model with a complex scalar field $S$ 
which is a $SU(2)_L$ singlet without $U(1)_Y$ charge:
${\cal L} = {\cal L}_{\rm kin}
           - {V}_{\Phi} 
           - {V}_{S}  
           - {V}_{\fsl{Z}_2}$,  
where ${\cal L}_{\rm kin}$ represents the kinetic term and
$V_{\Phi}^{}$ is the $Z_2$ symmetric part 
of the THDM potential with $m_3^2=0$. 
The potential $V_S$ for the complex scalar $S$ and the interaction term 
$V_{\fsl{Z}_2}$ between $S$ and $\Phi_{1,2}$ are given by  
${V}_{S} =  M_S^2 S^\dagger S
            + \kappa (S^\dagger S)^2
            + {V}_{Z_{2n}}^{}, \;\; M_S^2 > 0$,
with 

\vspace*{-2mm}
\noindent
\begin{eqnarray} 
 V_{Z_{2n}}^{} = \frac{\eta}{\Lambda^{2n-4}} 
         \left( S^{2n} + {\rm h.c.} \right), \;\; \eta \sim {\cal O}(1),
 \label{VM}
\end{eqnarray} 

\vspace*{-2mm}
\noindent
and 

\vspace*{-2mm}
\noindent
\begin{eqnarray} 
V_{\fsl{Z}_2}
            = \frac{\xi}{\Lambda^{2\ell-2}} 
             \left( S^{2\ell} \Phi_1^\dagger \Phi_2
            + {\rm h.c.} \right),\;\; \xi \sim {\cal O}(1), \label{VB}   
\end{eqnarray} 

\vspace*{-2mm}
\noindent
respectively. In Eqs.~(\ref{VM}) and (\ref{VB}), $\Lambda$ denotes
the cutoff scale of the model. 
We now set $n=1$ (case A) or 
$n=\ell$ (case B) with $\ell \geq 1$.
We note that $V_S$ has the $Z_{2n}$ symmetry under 
$S \to e^{i\frac{\pi}{n}} S$,
while $V_{\Phi}$ is $Z_2$ invariant under the transformation
$\Phi_1 \to - \Phi_1, \;\;\Phi_2 \to + \Phi_2$. 
The interaction term~(\ref{VB}) 
explicitly breaks both $Z_{2n}$ and $Z_2$. 
Some invariant terms under $Z_{2n}$ and $Z_2$ are not 
explicitly included here, as they are irrelevant 
to our conclusion. \\
\indent
Supposing that $M_S (\sim \Lambda)$ is much larger than the EWSB scale,
we integrate out the field $S$ and thereby obtain the THDM 
with the softly-broken $Z_2$ symmetry $(m_3^2 \ne 0)$
as the low-energy effective theory.
We estimate

\vspace*{-2mm}
\noindent
\begin{align}
& m_3^2 \sim \xi \eta^{\ell} \frac{1}{(4\pi)^{2\ell}} M_S^2, &
{\rm for \;\;\; Case \; A}, \\
& m_3^2 \sim \xi \eta \,\frac{1}{(4\pi)^{2(2\ell-1)}} M_S^2, &
{\rm for \;\;\; Case \; B}.
\end{align}

\vspace*{-2mm}
\noindent
If we take the cutoff $M_S = 4\pi v$ for 
Case~A or $M_S = (4\pi)^2 v$ for Case~B, 
we can obtain $m_3^2 \sim v^2/(4\pi)^2$ for $\ell=2$.

\vspace*{-2mm}
\section{Quark Mass Matrices}

Let us consider the extension of our model 
incorporating first two generation quarks. 
Under the discrete symmetry \cite{z2}, 
two types of Yukawa interactions are possible in the THDM, so called 
Model~I and Model II\cite{hhg}. 
Obviously Model I is inconsistent with our scenario, so that  
we here apply Model II. 
We then assume that $3 \times 3$ matrices $Y^{ij}_U$ and $Y^{ij}_D$ 
of the Yukawa coupling with up- and down-type quarks 
take the following forms,
$Y^{ij}_U \sim Y^{ij}_D \sim y,  \forall (i,j),   
\quad y \sim {\cal O}(1)$,
which lead to $m_t \gg m_c,m_u$ and $m_b \gg m_s,m_d$, and
the KM matrix becomes approximately diagonal.  
We can numerically reproduce the data for the mass spectrum 
and the KM matrix,
allowing small fluctuations of the Yukawa coupling constants.
Although we can avoid hierarchy among Yukawa couplings in this way,
subtle cancellation among the ${\cal O}(1)$ mass-matrix elements 
is required to obtain masses of light quarks.

It is possible to apply our scenario to the lepton sector. 
The tau lepton then receives the small mass
due to the similar mechanism to the bottom quark.
In this case, the Dirac mass of the tau neutrino could be 
produced around $m_t$.
To explain the tiny (Majorana) mass of the tau neutrino, 
additional mechanism such as the seesaw might be helpful. 

\vspace*{-2mm}
\section{Summary and Discussions}

We have proposed the mechanism to explain the mass hierarchy 
between the top and bottom quarks without fine tuning, 
starting from the vacuum with $(v_1,v_2)=(0,v)$.
Such a vacuum can exist when the $Z_2$ symmetry is exact.
The observed mass spectrum $m_t \gg m_b \ne 0$ is realized
via the small soft-breaking parameter $m_3^2$ for the $Z_2$ symmetry.
We have presented the model in which a small $m_3^2$ is
induced from the underlying physics above the cutoff scale of the
THDM. 
The size of $\tan\beta$ corresponds to the ratio $m_t/m_b \sim 40$.
The masses of the extra Higgs bosons $H$, $A$ and $H^\pm$ 
are expected to be ${\cal O}(v)$. 
In addition to the SM-like Higgs boson $h$, 
all the extra Higgs bosons in our model
are expected to be discovered at the CERN Large Hadron Collider (LHC).
Our prediction of $\sin (\beta-\alpha) \simeq 1$ can also be confirmed
at the LHC and Linear Collider's (LC's).
Our scenario may further be tested by measuring 
the $hhh$ coupling at future LC's\cite{hhh}.\\

\bibliographystyle{plain}

\begin{thebibliography}{1}
 \bibitem{hk} M.~Hashimoto, S.~Kanemura, hep-ph/0403005,
 Phys. Rev. D in press.

\bibitem{z2} S.~Glashow, S.~Weinberg, 
             \Journal{\PRD}{15}{1958}{1977}.  

\bibitem{cs} P.~Sikivie, et al., 
             \Journal{\NPB}{173}{189}{1980}.

\bibitem{gs_2hdm} H.E.~Haber, A.~Pomarol, 
              \Journal{\PLB}{302}{435}{1993};
              A.Pomarol and R.~Vega, 
              \Journal{\NPB}{413}{3}{1994}.

\bibitem{hhg} J.F.~Gunion, et al.,
{\it The Higgs Hunter's Guide}, Perseus Publishing, Cambridge, MA, 1990.

\bibitem{hhh} S.~Kanemura, et al.,
                  \Journal{\PLB}{558}{157}{2003}. 
                              
\end{thebibliography}

\end{document}